\documentclass[prl,aps,twocolumn,showpacs,preprintnumbers,amsmath,amssymb]{revtex4}
\usepackage{bm}
\usepackage{epsfig}

\begin{document}

\preprint{}

\title{Local moment, itinerancy and deviation from Fermi liquid behavior in Na$_x$CoO$_2$ for $0.71 \leq x \leq 0.84$}

\author{L. Balicas,$^1$ Y. J. Jo,$^1$ G. J. Shu,$^{2}$ F. C. Chou,$^{2,3}$ and P. A. Lee$^{4}$}

\affiliation{$^1$National High Magnetic Field Laboratory, Florida
State University, Tallahassee-FL 32310, USA}
\affiliation{$^2$Center for Condensed Matter Sciences, National
Taiwan University, Taipei 10617, Taiwan} \affiliation{$^3$National
Synchrotron Radiation Research Center, HsinChu 30076,Taiwan}
\affiliation{$^4$Department of Physics, Massachusetts Institute of
Technology, Cambridge, Massachusetts 02139, USA}

\date{\today}%
\begin{abstract}

Here we report the observation of Fermi surface (FS) pockets via
the Shubnikov de Haas effect in Na$_x$CoO$_2$ for $x = 0.71$ and
$0.84$, respectively. Our observations indicate that the FS
expected for each compound intersects their corresponding
Brillouin zones, as defined by the previously reported
superlattice structures, leading to small reconstructed FS
pockets, but only if a precise number of holes per unit cell is
\emph{localized}. For $0.71 \leq x < 0.75$ the coexistence of
itinerant carriers and localized $S =1/2$ spins on a paramagnetic
triangular superlattice leads at low temperatures to the
observation of a deviation from standard Fermi-liquid behavior in
the electrical transport and heat capacity properties, suggesting
the formation of some kind of quantum spin-liquid ground state.

\end{abstract}

 \maketitle

The Na cobaltates Na$_x$CoO$_2$ are the subject of an intense
interest due to i) an anomalously large thermoelectric response
for $0.7 < x <0.84$ \cite{ong, ong2}, ii) the discovery of
superconductivity upon hydration for $x=0.3$ \cite{takada}, and
iii) the overall complexity of their phase diagram as a function
of $x$ \cite{foo}. The thermoelectric response is particularly
intriguing, since it is claimed to be associated with the presence
of pronounced spin fluctuations \cite{ong} in the paramagnetic
region of the phase diagram where the magnetic susceptibility
displays Curie-Weiss (CW) behavior. CW response suggests the
existence of localized magnetic moments in these metallic systems.

Initially it was thought that small Na concentration $x$
corresponds to a lightly doped Mott insulator where correlation
effects should be strong.  However, it turns out that complicated
magnetic order an CW behavior usually associated with strong
correlations appear for $x \geq 0.5$. This is surprising because
the $x = 1$ limit is a band insulator with all Co in the $3+$
state with $S = 0$.  Compounds with $x$ slightly less than 1
should be thought of as hole doping of a band insulator, where
correlation effects are weak because dilute carriers are moving in
a featureless background.  It has been suggested that the
potential due to the ionized Na must play an important role
\cite{Merino} and calculations based on dynamical mean field
theory show that by increasing the Na potential CW behavior
appears \cite{Marianetti}. Just as in doping of semiconductors,
the ionized donor can trap carriers which form local moments once
Coulomb repulsion between carriers is taken into account.  This
point of view received strong support from the recent discovery of
well defined $\sqrt{13} \times \sqrt{13}$ and $\sqrt{12} \times
\sqrt{12}$ superlattice structures for $x = 0.84$ and $x = 0.71$,
respectively \cite{fangcheng}.  These superstructures can be
explained if the Na ions form clusters of di-vacancies for $x =
0.84$ and alternating layers of tri-vacancies and quadra-vacancies
for $x = 0.71$.  The tendency towards vacancy clustering was first
proposed by Roger {\em et al}., \cite{Roger} even though the
detailed realization turns out to be different from their original
proposal.  In any event, vacancy clustering favors binding of a
fraction of the carriers, because the clusters provide a stronger
and more dilute binding potential than individual Na ions.  This
led Chou {\em et al}. \cite{fangcheng} to propose a picture of
co-existing local moments sitting on the superlattice and
itinerant carriers for the doping range $x = 0.71$ to $x = 0.84$.

In order to find evidence for charge localization associated with
the superlattice structures, we studied the size of their
respective Fermi surfaces through electrical transport
measurements at very high magnetic fields and very low
temperatures to detect the Shubnikov the Haas (SdH) effect,
previously seen in samples with lower values of $x$ \cite{me}. All
crystals measured for this study, display typical resistivity
ratios $\rho$(300 K)$/\rho$(0.6 K) between 100 and 250 with
residual resistivities ranging from 5 to 10 $\mu \Omega$ cm (data
not shown here but data down to 4~K is reported in \cite{Shu}).
These values are nearly one order of magnitude lower than those
reported previously by other groups, see for example
\cite{ong2,li}. The exceptional quality of these crystals
motivated us to explore the low temperature properties of their
paramagnetic state. We performed a systematic study of the
electronic and heat capacity properties as function of $x$, with
$x$ having values $(0.71 \pm 0.01)$ $\leq x \leq $ $(0.747 \pm
0.01)$ as determined via electron micro analysis (EPMA).

Single crystals of Na$_{x}$CoO$_2$ were grown using the
floating-zone technique. Samples with different values of $x$ were
carefully produced via an electrochemical de-intercalation
procedure, for details about sample growth see Ref. \cite{Shu}.
Resistivity measurements down to $^3$He temperatures were
performed using standard four-terminal techniques by using either
a Physical Parameter Measurement System (PPMS) or a rotating
sample holder inserted into a $^3$He cryostat coupled to magnetic
fields up to $H = 45$ T provided by the hybrid magnet at the
National High Magnetic Field Lab. Heat capacity measurements via
the relaxation method down to $T = 350$ mK were also performed by
using the PPMS.
\begin{figure}[htb]
\begin{center}
\epsfig{file= 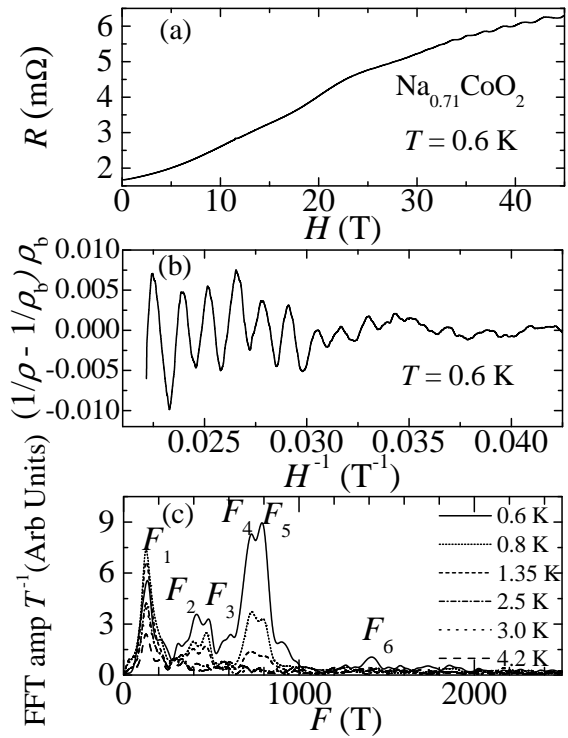, width = 7.4 cm} \caption {(a)
Resistance $R$ as a function of field $H$ for a Na$_{0.71}$CoO$_2$
single crystal at $T = 0.6$ K. (b) The SdH effect or the
oscillatory component of $R$ in (a) as a function of $H^{-1}$. (c)
Amplitude of the fast fourier transform normalized by temperature
$T$ for several values of $T$. At least six frequencies are
detected: $F_1 = 125$ T, $F_2 = 400$ T, $F_3 = 475$ T, $F_4 = 725$
T, $F_5 = 800$ T, and $F_6 = 1413$ T. The corresponding effective
masses in units of free electron mass are: $\mu_1 = 1.2 \pm 0.9$,
$\mu_2 = 6.9 \pm 1.6$, $\mu_3 = 4.4 \pm 0.4$, $\mu_4 = 12 \pm 1$,
$\mu_5 = 10 \pm 1.4$, respectively.}
\end{center}
\end{figure}

Figure 1 (a) Shows a typical raw resistance trace for a
Na$_x$CoO$_2$ single crystal with a Na concentration $x = 0.71 \pm
0.01$ as a function of field $H$ applied along the inter-plane
direction at a temperature $T$ = 0.6 K. The oscillatory component
seen at higher fields is the Shubnikov-de Haas (SdH) effect. We
define the SdH signal as $(1/ \rho - 1/ \rho_b) \rho_b$ (where
$\rho$ is the resistivity and $\rho_b$ is the background
resistivity obtained from a polynomial fit) and it is displayed in
Fig. 1(b) for the trace in (a). The fast Fourier transform (FFT)
of the SdH signal is shown in Fig. 1 (c) for several temperatures.
At least six peaks at well defined frequencies are observed in the
FFT spectra which through Onsager relation $F = A(\hbar/2 \pi e)$
correspond to cross sectional areas $A$ of the Fermi surface
perpendicular to the magnetic field. Two pairs $F_2$, $F_3$ and
$F_4$, $F_5$ may originate from splitting by interlayer hopping.
As in our previous reports \cite{me} the observation of several
Fermi surface pockets is at odds with all angle resolved
photoemission studies (ARPES) \cite{yang,arpes} which observe a
single hole-like Fermi surface sheet associated with the $A_{1g}$
band at the center of the hexagonal Brillouin zone. According to
ARPES the area of this sheet scales with the effective carrier
concentration $x^{\prime} = 1 - 2 A_{\mathrm{FS}}$, thus following
the Luttinger theorem \cite{yang}. For $x = 0.75$, ARPES indicates
a Fermi surface with an area corresponding to 12.5 \% of the
$A_{\mathrm{FBZ}}$  (area of the first Brillouin zone) when in
reality all the pockets detect by us have an $A_{\mathrm{FS}} < 3$
\% of $A_{\mathrm{FBZ}}$.  (For example, $F_2 = 400$~T corresponds
to $A_{\mathrm{FS}} \simeq 0.7$ \% of $ A_{\mathrm{FBZ}}$.)
\begin{figure}[htb]
\begin{center}
\epsfig{file= 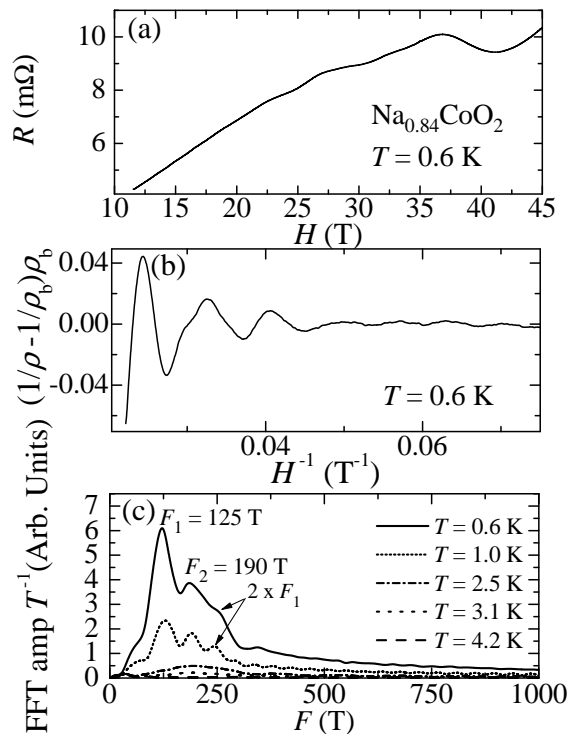, width = 7.4 cm} \caption {(a)
Resistance $R$ as a function of field $H$ for a Na$_{0.84}$CoO$_2$
single crystal at $T = 0.6 K$. (b) The oscillatory component of
$R$ in (a) as a function of $H^{-1}$. (c) Amplitude of the fast
fourier transform normalized by temperature $T$ for several values
of $T$. Two main frequencies are detected: $F_1 = 125$ T, $F_2 =
190$ T. The corresponding effective masses in units of free
electron mass are: $\mu_1 = 7 \pm 1$, $\mu_2 = 5.6 \pm 0.3$,
respectively.}
\end{center}
\end{figure}

Similar results are also obtained for $x = 0.84$. As in Fig. 1, in
Fig. 2 (a) we show a typical resistance trace for a
Na$_{0.84}$CoO$_2$ single crystal as a function of field at
$T=0.6$ K, in Fig. 2(b) the corresponding SdH signal, while Fig. 2
(c) shows the associated FFT spectrum. In contrast to $x=0.71$,
for $x =0.84$ one sees only two peaks at very low frequencies, and
related harmonics.

Fermi surface pockets are expected to result from the intersection
of the Fermi surface with the new Brillouin zones induced by the
Na superlattice structures seen for $x = 0.75$ and 0.84, namely
$\sqrt{12}a \times \sqrt{12}a$ and $\sqrt{13}a \times \sqrt{13}a$,
respectively \cite{fangcheng}.
\begin{figure}
\begin{center}
\epsfig{file=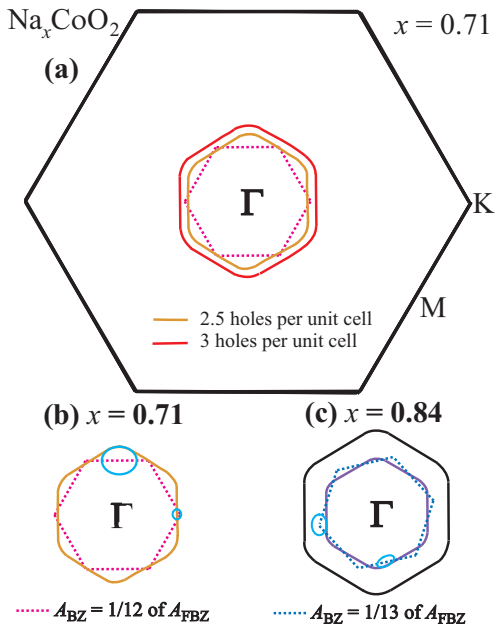, width=6.5 cm} \caption{(color
online)(a) The Fermi surface for $ x = 0.71 \simeq (1 - 3.5/12)$
within the hexagonal first Brillouin (FBZ) for 3 and 2.5 holes per
reconstructed unit cell \cite{fangcheng}, respectively in red and
in amber, assuming the absence of the $\epsilon^{\prime}_{g}$
pockets. The dotted hexagon depicts the superlattice induced new
Brillouin zone whose area is 1/12 of the area of the original FBZ.
b) Fermi surface pockets (in clear blue) can  result from the
intersection between the new Brillouin zone and the Fermi surface
depicted for 2.5 holes per reconstructed unit cell. c) Similarly,
for $x = 0.84 \simeq (1 - 2/13) $ the superlattice induced new
Brillouin zone (in blue), i.e., 1/13 of area of the original FBZ,
and the spin polarized Fermi surfaces for 2 and 1 hole per unit
formula, respectively in black and purple. Two Fermi surface
pockets only result from the intersection of the new Brillouin
zone and the originally depicted spin-polarized Fermi surface for
1 hole per reconstructed unit cell. Thus at least one hole per
reconstructed unit cell is \emph{localized}}
\end{center}
\end{figure}
In Figure 3 (a) we sketch the original (black line) and the
superlattice induced (magenta line) hexagonal Brillouin zones for
$x = 0.71$. Since $x = 0.71 \simeq 1 - 3.5/12$, one would naively
expect that $A_{\mathrm{FS}}$ would be given by 3.5 holes per
superlattice unit cell whose Brillouin zone has 1/12 of the area
of the original Brillouin zone. But as can be seen in figures 3
(a) and (b), when one takes into account the spin degeneracy, only
a Fermi surface whose area is given by at most 2.5 holes per
superlattice induced formula unit will end up intersecting the new
Brillouin zone leading to reconstructed Fermi surface pockets
(clear blue lines) of very different cross-sectional areas. Here,
following both ARPES \cite{yang,arpes} and our previous results
\cite{me}, we assumed the absence of the $\epsilon^{\prime}_{g}$
pockets. Thus the FS pockets seen by us are a very strong
indication that at least 1 hole per superlattice formula unit is
in fact \emph{localized} in Na$_{0.71}$CoO$_2$. Similarly, for $x
= 0.84 \simeq 1 - 2/13$ one expects to have a Fermi surface whose
area is given by 2 holes per unit cell which, in fact would
intersect the superlattice induced Brillouin zone. But A-type
antiferromagnetism, i.e., ferromagnetic planes coupled
antiferromagnetically, has been reported for $x = 0.84$
\cite{helme} what de facto lifts the spin degeneracy in this
system. Thus, as seen in Fig. 3 (c) a spin polarized Fermi
surface, i.e., a half-metallic state, conformed by \emph{one hole}
per formula unit would intersect the new Brillouin zone leading to
two small Fermi surface pockets as experimentally seen by us (Fig.
2 (c)). Consequently, also in this case we must have at least one
localized hole per formula unit. Furthermore, in our sketch there
are 2 pockets at the corners of the superlattice Brillouin zone,
and 3 at its sides. Since the total FS area for both types of
pockets should be equal, the respective ratio of their cross
sectional areas is 3 to 2, as precisely seen in our experiment.
Thus our results are clearly consistent with the presence of a
single Na di-vacancy per superlattice unit cell \cite{fangcheng}.

The interplay between localization and itinerancy, which in most
cases leads to rich new physical behavior, e.g. heavy-Fermions
\cite{stewart} and cuprate superconductors \cite{PALee}, is the
central theme in strongly correlated electron physics. Here, the
combination of localized but disordered $S = 1/2$ spins on a
frustrated triangular superlattice which in the paramagnetic state
interact with itinerant carriers, is expected to lead to
unconventional physical response, particularly ar low temperatures
where quantum spin fluctuations become predominant. In effect, as
we show below, the low temperature properties of the paramagnetic
state, i.e., $0.71 \leq x \leq 0.747$, are indeed quite anomalous.
\begin{figure}[htb]
\begin{center}
\epsfig{file=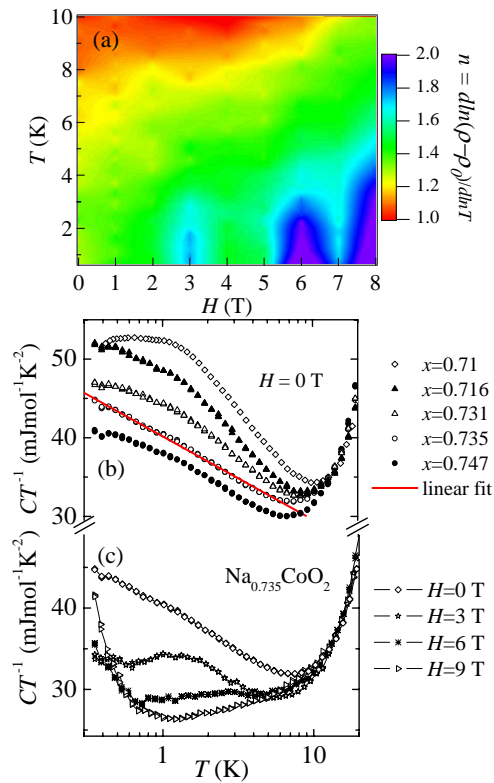, width = 7.8 cm} \caption {(color
online) (a) The evolution of the exponent of the resistivity $n =
\partial\ln{(\rho - \rho_0)} / \partial \ln{T}$ as a function of
field $H$ for $0.6 \leq T \leq 10$ K for a Na$_{0.735}$CoO$_2$
single crystal. (b) Heat capacity normalized by temperature $C/T$
as a function of $T$ for several Na concentrations $x$ in a
semi-log scale. Straight line represents a linear fit to the $x =
0.735$ data over the temperature range $0.35 \leq T \leq 7$ K. (b)
$C/T$ as a function of $T$ for $x=0.735$ and for several values of
the field $H$.}
\end{center}
\end{figure}
In Figure 4 (a) we present a contour plot displaying the evolution
of the exponent of the resistivity $n = \partial\ln{(\rho -
\rho_0)} / \partial \ln{T}$ for a Na$_{0.735}$CoO$_2$ single
crystal as a function of field $H$ for temperatures $T < 10$ K. As
is clearly seen, $n$ does not remain constant as function of
temperature for any field value. Furthermore, at $H = 0$ it
crossovers from two anomalous values, $n = 1$ to $n \sim 1.3$ at
the lowest temperatures. This anomalous exponent is consistent
with that found for the $x = 0.71$ samples \cite{Minhyea} and
disagrees with the exponent 2 found earlier in more disordered
samples \cite{li}. This value progressively increases with field
until it reaches the conventional Fermi liquid value $n=2$ for $H
> 5$ T. This anomalous behavior in the resistivity if followed by
a concomitant anomalous behavior in the heat capacity. In effect,
in Fig. 4 (b) we plot the heat capacity normalized by temperature
$C/T$ for several concentrations $0.71 \leq x \leq 0.747$ at zero
field. For $T < 10$ K, $C/T$ crossovers from a constant value, the
the electronic contribution in a Fermi liquid, at $x = 0.71$ to a
nearly logarithmic dependence on temperature for $0.71 < x \leq
0.747$. $C/T$ for $x=0.71$ shows a remarkably large Sommerfeld
coefficient $\gamma \simeq 52$ mJ/molK$^2$, which, to our
knowledge is the largest value ever reported for the Na$x$CoO$_2$
series. This value is nearly 60 \% higher than the one previously
reported for $x=0.72$ \cite{sales} and it probably reflects the
higher quality of our samples. Furthermore, as $x \rightarrow
0.75$ the nearly logarithmic dependence progressively spans over a
larger range in temperatures and eventually, as can be seen for
the $x = 0.735$ trace, the logarithmic dependence dominates $C/T$
over a decade in $T$ down to the lowest temperatures. The
application of a magnetic field suppresses this anomalous behavior
and a nearly constant but field dependent value of $C/T$ is
reached at lower temperatures prior to an upturn at the lowest
temperatures which results from the Co nuclear Schottky term
\cite{bruhwiler}, see Fig. 4(c).

In summary, deviation from standard Fermi-liquid behavior is
observed in the paramagnetic state of Na$_x$CoO$_2$ as $x
\rightarrow 0.75$, and as a result of the interplay between
itinerancy and localized $S=1/2$ moments on a triangular
superlattice. The upturn in the low temperature specific heat and
the anomalous scattering in the resistivity both point to the
existence of a novel kind of low energy excitations.  That these
excitations appear to be suppressed by a magnetic field suggests
that these are spin fluctuations.  One suggestion \cite{fangcheng}
is that the local moments form a spin liquid state \cite{anderson}
which co-exists with itinerant electrons.  While our data does not
allow us to draw any firm conclusion on the origin of the
anomalous behavior, it is clear that the cobaltates offer a new
window into possible novel ground states in a system with coupled
local moments and itinerancy.

The NHMFL is supported by NSF through NSF-DMR-0084173 and the
State of Florida. LB acknowledges the NHMFL in-house research
program. YJJ acknowledges support from the NHMFL-Schuller program.
FCC acknowledges support the National Science Council of Taiwan
under project number NSC-95-2112-M-002. PAL acknowledges DOE
number DE-FG02-03ER46076.

\end{document}